\documentclass[a4paper,12pt]{article}
\usepackage{amsmath,amssymb,cite,fancyhdr}

\newcommand{\al}{\alpha}
\newcommand{\be}{\beta}

\newcommand{\la}{\lambda}
\newcommand{\si}{\sigma}
\newcommand{\vep}{\varepsilon}

\newcommand{\De}{\Delta}

\newcommand{\bx}{{\boldsymbol x}}

\newcommand{\tf}{{\tilde f}}
\newcommand{\tK}{{\tilde K}}
\newcommand{\ttau}{{\tilde\tau}}

\newcommand{\cP}{{\mathcal P}}
\newcommand{\cR}{{\mathcal R}}

\newcommand{\RR}{{\mathbb R}}

\newcommand{\pa}{{\partial}}

\newcommand{\e}{{\rm e}}
\newcommand{\diff}{\operatorname{d}\!}

\def\ni{\noindent}
\def\ns{\normalsize}
\def\nem{\ns\em}
\numberwithin{equation}{section}
{

\oddsidemargin = .5cm
\textwidth = 15.5cm
\topmargin = 0cm

\begin{document}
\title{On the construction of evolution equations\\
admitting a master symmetry}
\author{\ns\textsc{F. Finkel\footnote{Corresponding author.
E-mail: federico@ciruelo.fis.ucm.es}}\\
\nem Departamento de F\'\i sica Te\'orica II\\
\nem Universidad Complutense\\
\nem 28040 Madrid\\
\nem Spain
\and\ns\textsc{A.S. Fokas}\\
\nem Department of Mathematics\\
\nem Imperial College\\
\nem London SW7 2BZ\\
\nem United Kingdom\vspace{.5cm}}

\date{\ns May 18, 2001}
\maketitle
\begin{abstract}
A method for constructing evolution equations
admitting a master symmetry is proposed.
Several examples illustrating the method are presented.
It is also noted that for certain evolution equations
master symmetries can be useful for obtaining new
conservation laws from a given one.
\end{abstract}
\vspace{5cm}
Mathematics Subject Classification (2000): 35Q53, 35Q58, 37K10\\[1mm]
Keywords: master symmetries, integrable evolution equations,
conservation laws \setcounter{footnote}{1} \thispagestyle{fancy}
\renewcommand{\headrulewidth}{0pt}
\rhead{To appear in {\em Physics Letters A}}
\newpage
\section{Introduction}
The concept of master symmetries was introduced in~\cite{FF81}.
This concept, which was further developed
in~\cite{CLL82,CLL83,Fu83,Fo87,Do93}, is deeply related to the so called
bispectral problem~\cite{ZM91}, as well as to Virasoro
algebras~\cite{GH96}. Among the direct applications of master
symmetries are the recursive construction of hierarchies of
infinitely many symmetries and of infinitely many gradients of
conserved densities for nonlinear integrable evolution equations.

In this paper we present an algorithm for constructing evolution
equations which are invariant under a given group of scaling
transformations and which possess a master symmetry. Moreover, as
a further indication of the importance of master symmetries, we
note that they can be useful for constructing new conservation
laws starting from a given one. We note that the method introduced
here can be applied to other types of equations such as the very
interesting class of non-linear equations recently introduced in~\cite{MS00}.

We first introduce the basic rings used throughout this paper.
Then we define symmetries, master symmetries, and scalings.\\

\ni{\bf Definition~1.1\/} {\em (The basic rings). Let $q(t,\bx)$,
$t\in\RR$, $\bx=(x_1,\dots,x_n)\in\RR^n$, be a smooth function of the
indicated variables. Let $\pa_{x_j}^{-1}$ denote the integral operators
$\pa_{x_j}^{-1}=\frac12\big(\int_{-\infty}^{x_j}-\int^{\infty}_{x_j}\big)$,
$j=1,\dots,n$. We shall denote by $\cR\{q\}$ the ring of
smooth functions of $q$ and its $x_j$- partial
derivatives; $\cR\{q,\pa^{-1}q\}$ will denote the ring of smooth
functions of $q$, its $x_j$- partial
derivatives, and its integrals $\pa_{x_j}^{-1}$.
If these smooth functions are polynomials, the corresponding
rings will be denoted by $\cP\{q\}$ and $\cP\{q,\pa^{-1}q\}$.
Similarly, if these smooth functions depend explicitly on
$\bx$ and on $t$ we will use the notation $\cR\{q,\pa^{-1}q,t,\bx\}$.}\\

\ni{\bf Definition~1.2\/} {\em (Symmetries). Let $\De(q)$ and
$\eta(q)$ belong to $\cR\{q,\pa^{-1}q,t,\bx\}$.
We call $\eta$ a\/ {\em symmetry} of the PDE
\begin{equation}\label{De}
\De(q)=0,
\end{equation}
iff
\begin{equation}\label{symm}
\De'\eta=0\qquad\text{when}\qquad\De=0,
\end{equation}
where $\De'$ denotes the Fr\'echet derivative of $\De$ with
respect to q, i.e.,
\begin{equation}\label{frechet}
  \De'(q)\eta(q)=\frac{\pa}{\pa\vep}\,\De(q+\vep\eta)\big|_{\vep=0}.
\end{equation}}

\subsection*{Constructing master symmetries}

Let $q$ satisfy the evolution PDE
\begin{equation}\label{evPDE}
  q_t=K_1(q)\,,\qquad K_1(q)\in\cR\{q,\pa^{-1}q\}\,.
\end{equation}
The definition of a symmetry implies that the function
$K_2(q)\in\cR\{q,\pa^{-1}q\}$ is a symmetry of this evolution PDE iff
\begin{equation}\label{K1K2}
  \big[K_1(q),K_2(q)\big]_L=0\,,
\end{equation}
where the Lie commutator $[\,,]_L$ is defined by
$$
\big[A(q),B(q)\big]_L=A'B-B'A\,.
$$

We note that since $K_1$ and $K_2$ are $x_j$-independent they
commute with $q_{x_j}$, i.e.
\begin{equation}\label{qK1qK2}
  [q_{x_j},K_1]_L=[q_{x_j},K_2]_L=0\,.
\end{equation}
Thus for evolution equations the question of constructing
symmetries which are independent of $x_j$ and of $t$,
reduces to the question of constructing an Abelian Lie algebra with elements
$\{q_{x_j},K_1,K_2\}$. Such a construction can be carried out
using a master symmetry.\\

\ni{\bf Definition 1.3\/} {\em (Master symmetries). Suppose that
$K_1(q),K_2(q)\in\cR\{q,\pa^{-1}q\}$ commute, i.e., suppose that they
satisfy Eq.~\eqref{K1K2}. We call $\tau(q)\in\cR\{q,\pa^{-1}q,\bx\}$
a\/ {\em master symmetry} associated with $q_{x_j}$, $K_1$, and $K_2$, iff
\begin{equation}\label{master}
  [q_{x_j},\tau]_L=c_1K_1\,,\qquad [K_1,\tau]_L=c_2K_2\,,
\end{equation}
where $c_1$ and $c_2$ are constants.}\\

This definition implies that the Abelian algebra generated by
$\{q_{x_j},K_1,K_2\}$ is constructed solely from a master symmetry.

The existence of a master symmetry implies the existence of a
generalized symmetry $K_2$.\footnote{It is called generalized in
juxtaposition to Lie point because either it depends
nonlinearly on first-order derivatives or it involves
higher-order derivatives, see~\cite{Fo87} for further discussion.}
Although in all known examples the recursive use of $\tau$,
$$
[K_j,\tau]_L=c_j K_{j+1},
$$
generates infinitely many symmetries, their existence does not
follow from Definition~1.3. However, it was conjectured
in~\cite{Fo80} that the existence of one generalized symmetry
implies infinitely many symmetries. This conjecture was shown to
be true in~\cite{SW98} for a large class of evolution equations in
one space dimension. This provides further support for the
usefulness of constructing master symmetries, since they guarantee
the existence of at least one generalized symmetry. In what
follows we present such a construction based on the notion of
scaling:\\

\ni{\bf Definition~1.4\/} {\em (Scalings). Suppose
that $K(q)\in\cR\{q,\pa^{-1}q,\bx\}$ is invariant up to a
multiplicative constant under the
transformation $q\mapsto\al^\la q$, $x_j\mapsto\al^{\la_j} x_j$,
i.e.
\begin{equation}\label{scaling}
  K\big(\al^\la q(t,\al^{\la_1}x_1,\ldots,\al^{\la_n}x_n)\big)
  =\mu K\big(q(t,x_1,\dots,x_n)\big),
\end{equation}
where $\al$, $\la$, $\la_j$, $\mu$ are constants. Then we say that
$K(q)$ admits the\/ {\em scaling} $\tau_0(q)$, where $\tau_0$ is defined
by
\begin{equation}\label{tau0}
  \tau_0(q)=\la q-\sum_1^n\la_j x_j q_{x_j}\,.
\end{equation}
If the constant $\mu$ equals one, we say that $K(q)$ is
an\/ {\em absolute invariant} of the scaling $\tau_0$.}\\

It is shown in~\cite{Fu83} that if $K(q)\in\cR\{q,\pa^{-1}q,\bx\}$
admits the scaling $\tau_0(q)$, then
\begin{equation}\label{LieKtau0}
  [K,\tau_0]_L=cK,
\end{equation}
where $c$ is a constant.\\

\ni{\bf Observation~1.1\/} {\em (Candidates for master symmetries).
Suppose that $K_1(q)\in\cR\{q,\pa^{-1}q\}$ admits
the scaling $\tau_0(q)$ defined by Eq.~\eqref{tau0}.
Assume that there exists a function $f(q)\in\cR\{q,\pa^{-1}q,\bx\}$
such that $\tau_0$ can be written in the form
\begin{equation}\label{obs13}
  \tau_0=[q_{x_j},f]_L\,.
\end{equation}
Define $\tau(q)\in\cR\{q,\pa^{-1}q,\bx\}$ by
\begin{equation}\label{tau}
  \tau=[K_1,f]_L+\tau_1\,,
\end{equation}
where
$\tau_1(q)\in\cR\{q,\pa^{-1}q,x_1,\dots,x_{j-1},x_{j+1},\dots,x_n\}$
is an arbitrary $x_j$-independent function. Then $\tau$ satisfies
\begin{equation}\label{qxjtau}
[q_{x_j},\tau]_L=c_1K_1\,.
\end{equation}}

Indeed, since $\tau_1$ is $x_j$-independent, it commutes with
$q_{x_j}$; thus
\begin{equation}\label{proof1.1}
[q_{x_j},\tau]_L=\big[q_{x_j},[K_1,f]_L\big]_L=
\big[K_1,[q_{x_j},f]_L\big]_L-\big[f,[q_{x_j},K_1]_L\big]_L\,,
\end{equation}
where we have used the fact that $[\,,]_L$ satisfies the Jacobi
identity. Since $K_1$ is $x_k$-independent for all $k$ and admits
the scaling $\tau_0$, Eq.~\eqref{proof1.1} becomes
$$
  [q_{x_j},\tau]_L=[K_1,\tau_0]_L=c_1 K_1\,.
$$

The above observation suggests the following algorithm for
constructing equations which admit a given scaling $\tau_0$ and a
master symmetry:\\

\ni{\bf Construction~1.1\/} {\em (Construction of master symmetries).\\
1. Start with a general function $K_1(q)\in\cR\{q,\pa^{-1}q\}$
which remains invariant up to a multiplicative constant
under the transformation $q\mapsto\al^\la q$, $x_j\mapsto\al^{\la_j} x_j$
(see Eq.~\eqref{scaling}). Let the associated scaling $\tau_0$
be defined by Eq.~\eqref{tau0}. The function $K_1(q)$ involves some
absolute invariants $b_j(q)\in\cR\{q,\pa^{-1}q\}$ of the scaling $\tau_0$.\\
2. Find a function $f(q)\in\cR\{q,\pa^{-1}q,\bx\}$ such that Eq.~\eqref{obs13} is
satisfied.\\
3. Define $\tau(q)\in\cR\{q,\pa^{-1}q,\bx\}$ by Eq.~\eqref{tau} and $K_2(q)$ by
\begin{equation}\label{K2alg}
  K_2=[K_1,\tau]_L\,.
\end{equation}
If there exist absolute invariants $b_j(q)$ and a function $\tau_1(q)$
such that $K_2\in\cR\{q,\pa^{-1}q\}$ and
$[K_1,K_2]_L=0$, then $\tau$ is a master symmetry
associated with $q_{x_j}$, $K_1$, $K_2$.}\\

We note that the requirement $K_2\in\cR\{q,\pa^{-1}q\}$ is
automatically satisfied in one space dimension ($n=1$). Indeed,
$$
[q_x,K_2]_L=\big[q_x,[K_1,\tau]_L\big]_L=
\big[K_1,[q_x,\tau]_L\big]_L-\big[\tau,[q_x,K_1]_L\big]_L=0\,,
$$
where we have used Eq.~\eqref{qxjtau} and the fact that $K_1$
commutes with $q_x$.\\

\ni{\bf Example 1.1\/} The most general third-order function
$K_1(q)\in\cP\{q\}$ which remains invariant up to a multiplicative constant
under the scaling
\begin{equation}\label{scalingex11}
q\mapsto\al^{-1} q,\qquad x\mapsto\al^2 x
\end{equation}
is given by
\begin{equation}\label{K1ex11}
  K_1=q_3+3 q^2 q_2+\be_1 q q_1^2+\be_2 q^4 q_1+\be_3 q^7\,,
\end{equation}
where throughout the paper
$$
q_j=\pa_x^jq\,,
$$
and $\be_1,\be_2,\be_3$ are constants.\footnote{Note that any absolute invariant
of the scaling~\eqref{scalingex11} in $\cP\{q\}$ is necessarily
a constant.} The scaling~\eqref{scalingex11} implies that
$\tau_0=q+2xq_x$. The class of solutions of the equation
$\tau_0=[q_x,f]_L$ depend on the form of $f$. If $f\in\cR\{q,\pa^{-1}q,x\}$,
then
$$
f=xq+x^2q_1+g(q)\,,
$$
where $g$ is an arbitrary function in $\cR\{q,\pa^{-1}q\}$. We take $g=0$,
define $\tau$ by Eq.~\eqref{tau} with $\tau_1=0$,
and determine $K_2$ from Eq.~\eqref{K2alg}. A straightforward
computation shows that $K_2$ commutes with $K_1$ if and only if
$$
\be_1=9\,,\qquad \be_2=3\,,\qquad \be_3=0\,.
$$
We recover in this way the Ibragimov--Shabat
equation~\cite{IS81,Ca87}
\begin{equation}\label{K1ex11final}
  q_t=q_3+3 q^2 q_2+9 q q_1^2+3 q^4 q_1\,;
\end{equation}
this equation admits the master symmetry
\begin{equation}\label{MSex11}
\tau=2x\big(q_3+3 q^2 q_2+9 q q_1^2+3 q^4 q_1\big)+3q_2+10q^2q_1+q^5\,.
\end{equation}

We note that the master symmetry of Burgers equation, both master
symmetries of the Kadomtsev--Petviashvili (KP) equation, and the
master symmetries of several other equations (see
Eqs.~\eqref{tauburg},~\eqref{tauKP},~\eqref{ttauKP},~\eqref{finaltauex22},
\eqref{tauSK}) can be obtained through Construction~1.1.

\subsection*{Master symmetries and conservation laws}

It is well known~\cite{Ol93} that certain symmetries can be used
to construct a new conservation law from a given one. However,
not all symmetries can be used for this purpose since some
symmetries yield trivial conservation laws. For example, the
Korteweg--de Vries (KdV) equation
\begin{equation}\label{potKdV}
  \pa_t q=\pa_x(q_2+3q^2)\,,
\end{equation}
is in a form of a conservation law and also it admits
infinitely many symmetries $K_j\in\cP\{q\}$.
If one starts with the conservation law~\eqref{potKdV}, all these
symmetries yield a trivial conservation law.

It will be shown in Section~3 that starting from~\eqref{potKdV},
the time-dependent generalized symmetry associated with the
master symmetry of the KdV equation can be used to generate
a hierarchy of nontrivial conservation laws.

\section{Constructing master symmetries}

\ni{\bf Example~2.1\/} The most general second-order function
$K_1\in\cP\{q\}$ which remains invariant
up to a multiplicative constant under the scaling
\begin{equation}\label{scalingex21}
q\mapsto\al^{-1} q,\qquad x\mapsto\al\,x
\end{equation}
is given by
\begin{equation}\label{burgersbeta}
  K_1=q_2+2qq_1+\be q^3\,.
\end{equation}
The above scaling implies that
\begin{equation}\label{tau0burg}
\tau_0=q+xq_1\,.
\end{equation}
The general solution of Eq.~\eqref{obs13} in $\cR\{q,\pa^{-1}q,x\}$ is
given by
$$
f=xq+\frac{x^2}2\,q_1+g(q)\,,
$$
where $g$ is an arbitrary function in $\cR\{q,\pa^{-1}q\}$.
Define $\tau$ by Eq.~\eqref{tau} with $g=\tau_1=0$, i.e.,
\begin{equation}\label{tauburgbeta}
\tau=[K_1\,,f]_L=2xK_1+3q_1+2q^2\,.
\end{equation}
A simple computation shows that the function $K_2$ defined by
\begin{equation}\label{K2burg}
K_2=[K_1\,,\tau]_L=4\Big(q_3+3qq_2+3q_1^2+(3+\be)\,q^2q_1+\frac{3\be}2\,q^4\Big)
\end{equation}
commutes with $K_1$ iff $\be=0$. In summary, the Burgers equation
\begin{equation}\label{burgers}
  q_t=q_2+2qq_1
\end{equation}
admits the master symmetry
\begin{equation}\label{tauburg}
\tau=2x(q_2+2qq_1)+3q_1+2q^2\,.
\end{equation}\vskip 1mm

\ni{\bf Example~2.2\/} Consider the function $K_1\in\cR\{q\}$
given by
\begin{equation}\label{K1ex22}
  K_1=q_3+3b_1(q)\,\frac{q_2^2}{q_1}+3b_2(q)\,q_1q_2+b_3(q)q_1^3\,,
\end{equation}
where $b_j(q)$ are arbitrary smooth functions of $q$. We assume that $b_1(q)$
is not identically zero. The function $K_1$ remains invariant up to a
multiplicative constant under the scaling $q\mapsto q$, $x\mapsto\al\,x$;
this corresponds to the function $\tau_0=x q_1$. The general solution of
Eq.~\eqref{obs13} in $\cR\{q,\pa^{-1}q,x\}$ is of the form
$$
f=\frac{x^2}2\,q_1+g(q)\,,
$$
where $g\in\cR\{q,\pa^{-1}q\}$. Define $\tau$ by Eq.~\eqref{tau}
with $g=\tau_1=0$, i.e.,
\begin{equation}\label{tauex22}
\tau=[K_1\,,f]_L=3\big(xK_1+(1+2b_1)q_2+b_2 q_1^2\big)\,.
\end{equation}
A straightforward computation shows that the function $K_2$
defined by Eq.~\eqref{K2alg} commutes with $K_1$ iff
$$
b_1=-\frac 14\,,\qquad b_2=b(q)\,,\qquad b_3=b(q)^2+2b'(q)\,,
$$
where $b(q)$ is an arbitrary smooth function of $q$,
and $b'(q)=\diff b/\!\diff q$. In summary, the evolution equation
\begin{equation}\label{eqex22}
q_t=q_3-\frac{3q_2^2}{4q_1}+3b\,q_1q_2+(b^2+2b')q_1^3
\end{equation}
admits the master symmetry
\begin{equation}\label{finaltauex22}
\tau=x\Big(q_3-\frac{3q_2^2}{4q_1}+3b\,q_1q_2+(b^2+2b')q_1^3\Big)
+\frac12\,q_2+b\,q_1^2\,.
\end{equation}
We note that~\eqref{eqex22} can be obtained from the equation~\cite{MSS91}
$$
u_t=u_3-\frac{3u_2^2}{4u_1}
$$
by the transformation
$$
u=\be_1\int^q\e^{2\int^s b(r)\diff r}\diff s+\be_2\,,
$$
where $\be_1$ and $\be_2$ are constants.\\

\ni{\bf Example~2.3\/} Consider the KP equation in the variable
$q(t,x,y)$,
\begin{equation}\label{KP}
  q_t=K_1\,,\qquad K_1=q_{xxx}+6qq_x+3\pa_x^{-1}q_{yy}.
\end{equation}
It is well-known that the KP equation admits infinitely
many symmetries and conservation laws.
The function $K_1$ defining the KP equation admits the scaling
\begin{equation}\label{tau0KP}
\tau_0=2q+xq_x+2yq_y\,.
\end{equation}
It may be easily verified that the function $f\in\cR\{q,\bx\}$ given by
$$
f=y(2q+xq_x+yq_y)
$$
satisfies Eq.~\eqref{obs13} for $q_y$, i.e.,
$$
\tau_0=[q_y\,,f]_L.
$$
The function $\tau\in\cR\{q,\pa^{-1}q,\bx\}$ defined by Eq.~\eqref{tau}
with $\tau_1=0$,
\begin{equation}\label{tauKP}
\tau=[K_1\,,f]_L=3\big(yK_1+2xq_y+4\pa_x^{-1}q_y\big)\,,
\end{equation}
is a master symmetry of the KP equation~\cite{OF82,CLL83}.
Indeed, the function $K_2\in\cP\{q,\pa^{-1}q\}$ defined by
\begin{equation}\label{K2KP}
K_2=[K_1\,,\tau]_L=
36\,\big(q_{xxy}+4qq_y+2q_x\pa_x^{-1}q_y+\pa_x^{-2}q_{yyy}\big)
\end{equation}
commutes with $K_1$.

The alternative equation
\begin{equation}\label{altobs13}
\tau_0=[q_x\,,\tf]_L
\end{equation}
also leads to a master symmetry of the KP equation. Consider the
solution of this equation given by
$$
\tf=x\Big(2q+\frac12\,xq_x+2yq_y\Big)\,.
$$
Define $\ttau_1\in\cR\{q,\pa^{-1}q,y\}$ by
$$
\ttau_1=6y\big(q_{xxy}+6qq_y+4q_x\pa_x^{-1}q_y+3\pa_x^{-2}q_{yyy}\big)
+3\big(q_{xx}+4q^2+2q_x\pa_x^{-1}q+11\pa_x^{-2}q_{yy}\big)\,.
$$
Then the function $\ttau\in\cR\{q,\pa^{-1}q,\bx\}$
defined by Eq.~\eqref{tau}, i.e.,
\begin{equation}\label{ttauKP}
\ttau=[K_1\,,\tf]_L+\ttau_1=3xK_1+\frac13\,yK_2
+6\big(2q_{xx}+4q^2+q_x\pa_x^{-1}q+3\pa_x^{-2}q_{yy}\big)
\end{equation}
is a master symmetry of the KP
equation~\cite{CLL83}.\footnote{Notice that $\ttau$ reduces to
the master symmetry of the KdV equation if $q$ does not depend on
$y$.} Indeed, it may verified that the function
\begin{align*}
\tK_2=[K_1\,,\ttau]_L=9\big(&q_{xxxxx}+10 q q_{xxx}+20 q_xq_{xx}
+10 q_{xyy}+30q^2q_x+10q_x\pa_x^{-2}q_{yy}\\[1mm]
&+20q_y\pa_x^{-1}q_y+20q\pa_x^{-1}q_{yy}+10\pa_x^{-1}(qq_{yy})
+10\pa_x^{-1}q_y^2+5\pa_x^{-3}q_{yyyy}\big)
\end{align*}
commutes with $K_1$. We note that
$$
K_3=[K_2\,,\tau]_L=12\tK_2\,.
$$
In general, the master symmetry $\ttau$ generates
the odd time-independent generalized symmetries $K_{2j+1}$
of the KP hierarchy $K_{j+1}=[K_j\,,\tau]_L$.\\

\ni{\bf Example~2.4\/} The function $K_1\in\cP\{q\}$ given by
\begin{equation}\label{K1SK}
  K_1=q_7+7qq_5+14q_1q_4+21q_2q_3+14q^2q_3+42qq_1q_2+7q_1^3
      +\frac{28}3\,q^3q_1
\end{equation}
is the first time-independent higher-order symmetry of the Sawada--Kotera (SK)
equation~\cite{SK74}
\begin{equation}\label{SK}
  q_t=\tK_1\,,\qquad \tK_1=q_5+5qq_3+5q_1q_2+5q^2q_1\,.
\end{equation}
The function $K_1$ in Eq.~\eqref{K1SK} admits the scaling
$$
\tau_0=q+\frac12\,x\,q_1\,.
$$
A straightforward computation shows that the function $f\in\cR\{q,x\}$
given by
$$
f=x\,q+\frac14\,x^2q_1
$$
solves Eq.~\eqref{obs13}. Define $\tau_1\in\cR\{q,\pa^{-1}q\}$ by
$$
\frac27\,\tau_1=3q_6+\tK_1\pa^{-1}_x q
+22qq_4+32q_1q_3+17q_2^2+38q^2q_2
+43qq_1^2-5q_1\pa^{-1}_x q_1^2
+\frac53\,q_1\pa^{-1}_x q^3+\frac{16}3\,q^4.
$$
Then the function $\tau\in\cR\{q,\pa^{-1}q,x\}$ determined
by Eq.~\eqref{tau}, i.e.,
\begin{equation}\label{tauSK}
\begin{aligned}
\tau=[K_1\,,f]_L+\tau_1=\frac72\Big(&xK_1+\tK_1\pa^{-1}_x q+8q_6+46qq_4
+75q_1q_3+44q_2^2+68q^2q_2\\
&+79qq_1^2-5q_1\pa_x^{-1}q_1^2+\frac53\,q_1\pa_x^{-1}q^3+8q^4\Big),
\end{aligned}
\end{equation}
is a master symmetry associated with $q_1$, $K_1$, and
$K_2=[K_1\,,\tau]_L$. In fact, the master symmetry $\tau$ generates
two hierarchies of infinitely many symmetries of the SK
equation~\cite{FO82}, namely
$$
K_{j+1}=[K_j\,,\tau]_L\,,\qquad
\tK_{j+1}=[\tK_j\,,\tau]_L\,,\qquad
j\geq 1.
$$

\section{Master symmetries and conservation laws}

In this section we shall show that for certain equations
the master symmetries can be useful
for constructing new conservation laws from a given
one. We first formally define conservation laws and trivial
conservation laws. Then we recall a well-known result~\cite{Ol93}
on the action of a symmetry on a conservation law.\\

\ni{\bf Definition~3.1\/} {\em (Conservation laws). We say that
Eq.~\eqref{De} admits the\/ {\em conservation law} characterized
by $\{T_j(q)\}_0^n$, where $T_j\in\cR\{q,\pa^{-1}q,\bx\}$, $j=0,\dots,n$, iff
\begin{equation}\label{cl}
  D_t T_0+\sum_{j=1}^n D_j T_j=0\qquad\text{when}\qquad\De=0,
\end{equation}
where $D_t$ and $D_j$ denote total differentiation with respect to
$t$ and $x_j$, i.e.,
\begin{equation}\label{DtDj}
  D_t=\frac\pa{\pa t}+q_t\frac\pa{\pa q}+\cdots\,,\qquad
  D_j=\frac\pa{\pa x_j}+q_{x_j}\frac\pa{\pa q}+\cdots\,.
\end{equation}
The component $T_0(q)$ of a conservation law is called the\/ {\em
conserved density}. The vector formed by the remaining components
$\big(T_1(q),\dots,T_n(q)\big)$ is called the\/ {\em flux}. A
conservation law is\/ {\em local} if $T_j(q)\in\cR\{q,\bx\}$,
$j=0,\dots,n$.}\\

\ni{\bf Definition~3.2\/} {\em (Trivial conservation laws). A
conservation law is\/ {\em trivial} if there exist smooth
functions $Q_{jk}\in\cR\{q,\pa^{-1}q,\bx\}$, $j,k=0,\dots,n$, such that
$Q_{jk}=-Q_{kj}$ and
\begin{equation}\label{trivial}
  T_j=D_tQ_{j0}+\sum_{k=1}^{n}D_kQ_{jk},\quad j=0,\dots,n\,,
  \qquad\text{when}\qquad\De=0\,.
\end{equation}
A trivial conservation law will be denoted as $\{T_j\}_0^n\sim 0$.
Two conservation laws are\/ {\em equivalent} if they differ by a
trivial conservation law.}\\

The Definitions~1.2 and 3.1 of symmetries and conservation laws imply:\\

\ni{\bf Observation~3.1\/} {\em (Symmetries and conservation laws).
Assume that the PDE~\eqref{De} possesses a symmetry
$\eta(q)\in\cR\{q,\pa^{-1}q,\bx\}$ and
a conservation law characterized by $\{T_j(q)\}_0^n$,
$T_j\in\cR\{q,\pa^{-1}q,\bx\}$,
see Definitions~1.2 and~3.1. Then this equation also possesses a
conservation law characterized by $\{T'_j(q)\eta(q)\}_0^n$.}

Indeed, it follows from Definitions~1.2 and~3.1 that
\begin{equation}\label{obs1}
  \frac\pa{\pa\vep}\,D_t T_0(q+\vep\eta)\Big|_{\vep=0}+
  \frac\pa{\pa\vep}\,\sum_{j=1}^n
  D_jT_j(q+\vep\eta)\Big|_{\vep=0}=0,
\end{equation}
which implies the desired result.

This observation provides an effective way to compute conserved
densities for self-adjoint linear systems~\cite{Ol93}.\\

\ni{\bf Example~3.1\/} Let $q$ satisfy,
\begin{equation}\label{wave}
  q_{tt}=q_{xx}+q_{yy}\,.
\end{equation}
This equation admits the conservation law characterized by
$$
T_0(q)=q_tq_x\,,\qquad
T_1(q)=-qq_{yy}-\frac12\,(q_t^2+q_x^2+q_y^2)\,,\qquad
T_2(q)=qq_{xy}\,.
$$
Consider the Lorentz-boost symmetry $\eta(q)=xq_t+tq_x$. Then
\begin{equation}\label{clwave}
\big\{T'_j\eta\big\}_0^2
\sim\Big\{xq_xq_{yy}+\frac12\,\big(q_t^2+q_x^2\big)\,,
    -q_t(q_x+xq_{yy})\,,
    x(q_tq_{xy}-q_xq_{ty})\Big\}
\end{equation}
is also a conservation law of equation~\eqref{wave}.
We have subtracted here the trivial conservation law determined
by (see Eq.~\eqref{trivial})
$$
Q_{01}=tq_tq_x+\frac x2\,\big(q_t^2+q_x^2\big)\,,\qquad
Q_{02}=0\,,\qquad
Q_{12}=-q(tq_{xy}+xq_{ty})\,.
$$\smallskip

We note that if $T_j(q)$, $j=0,\dots,n$, are invariant with
respect to the symmetry $\eta(q)$ under consideration
then $\{T'_j\eta\}_0^n\sim 0$, thus
the Observation~3.1 does not yield a new conservation law.\\

\ni{\bf Example~3.2\/} Consider the KdV equation~\eqref{potKdV}.
The KdV equation admits infinitely many symmetries
$K_j\in\cP\{q\}$ and local conservation laws.
The KdV equation is in a form of a conservation law, where
\begin{equation}\label{T0T1KdV}
  T_0(q)=q\,,\qquad T_1(q)=-q_2-3q^2\,.
\end{equation}
However, since $\{T'_0K_j,T'_1K_j\}\sim 0$ none of the symmetries $K_j$
yield new (nontrivial) conservation laws.

The KdV equation also possesses $x,t$-dependent symmetries
$\si_j(q)\in\cR\{q,\pa^{-1}q,t,x\}$. These symmetries are
intimately related to the master symmetry, see the Appendix.
In particular, a time-dependent symmetry of the KdV equation
is given by
\begin{equation}\label{KdVxtsymm}
\si=tK_2+\tau\,,
\end{equation}
where
$$
\tau=\frac 13\,\big(x K_1+4 q_2+8q^2+2q_1\pa_x^{-1}q\big)
$$
is the master symmetry of the KdV equation and
$$
K_2=[K_1\,,\tau]_L=q_5+10qq_3+20q_1q_2+30q^2q_1
$$
is its first generalized time-independent symmetry.
The symmetry $\si$ gives rise to the conservation law
\begin{equation}\label{T10T11KdV}
  T_0^{(1)}(q)=q^2\,,\qquad T_1^{(1)}(q)=-2qq_2+q_1^2-4q^3\,.
\end{equation}
Indeed, on solutions of the KdV equation we have
$$
\big\{T'_0\si\,,T'_1\si\big\}=\big\{T_0^{(1)},T_1^{(1)}\big\}
+\big\{D_xQ\,,-D_tQ\big\}\,,
$$
where
$$
Q=t\big(q_4+10 q q_2+5 q_1^2+10
q^3\big)+x\Big(\frac13\,q_2+q^2\Big)+q_1+\frac23\,q\pa_x^{-1}q\,.
$$
It should emphasized that the conserved density $T_0^{(1)}$
of the conservation law~\eqref{T10T11KdV}
can be generated solely by $\tau$, up to a total $x$-derivative.
The recursive application of the Fr\'echet derivative with respect to
$\si$ can be used to obtain the well-known sequence of
conservation laws associated with the KdV equation:
\begin{align*}
& \big\{T_0^{(2)},T_1^{(2)}\big\}=\big\{\!-\!q_1^2+2q^3\,,
    2q_1q_3-q_2^2-6q^2q_2+12qq_1^2-9q^4\big\},\\
& \big\{T_0^{(3)},T_1^{(3)}\big\}=\big\{q_2^2-10qq_1^2+5q^4\,,
    -2q_2q_4+q_3^2+20qq_1q_3-16qq_2^2-10q_1^2q_2-20q^3q_2\\
& \qquad\qquad\quad\, +90q^2q_1^2-24q^5\big\},\\
& \big\{T_0^{(4)},T_1^{(4)}\big\}=\big\{\!-\!q_3^2+14qq_2^2-70q^2q_1^2+14q^5\,,
    2q_3q_5-q_4^2-28qq_2q_4+20qq_3^2+28q_1q_2q_3\\
& \qquad\qquad\quad\, +140q^2q_1q_3-2q_2^3-154q^2q_2^2
    -140qq_1^2q_2-70q^4q_2+35q_1^4+560q^3q_1^2-70q^6\big\},
\end{align*}
where we have omitted an irrelevant global constant and a trivial
conservation law. The time-dependent symmetry $\si$ thus plays
for conservation laws the same role as the
adjoint of the recursion operator for their {\em conserved gradients};
see Ref.~\cite{Fo87}. The formalism based on the time-dependent
symmetry $\si$ associated with the master symmetry is
more convenient, because it provides explicit expressions for the
conserved densities $T_0^{(k)}$, and what is more important,
for the corresponding fluxes $T_1^{(k)}$.\\

\ni{\bf Remark~3.1\/} In some cases, the time-dependent
symmetry associated with the master symmetry yields
trivial conservation laws. For example, the
Burgers equation~\eqref{burgers}
is also in the form of a conservation law, with
\begin{equation}\label{burgCL}
  T_0(q)=q\,,\qquad T_1(q)=-q_1-q^2\,.
\end{equation}
Unlike the KdV equation, the Burgers equation is diffusive thus
does not possess infinitely many conservation laws.
In this case $\si=tK_2+\tau$, where $\tau$ and $K_2$ are respectively
given by Eqs.~\eqref{tauburg} and~\eqref{K2burg} with $\be=0$,
yields a trivial conservation law, namely
$$
\big\{T'_0\si\,,T'_1\si\big\}=\big\{D_xQ\,,-D_tQ\big\}\,,
$$
where
$$
Q=4t\big(q_2+3qq_1+q^3\big)+2x\big(q_1+q^2\big)+q\,.
$$
The situation with Eq.~\eqref{K1ex11final} is similar to the Burgers equation.

\section*{Acknowledgements}

F.F. would like to express his gratitude to
the Department of Mathematics of Imperial College for its hospitality.
A.S.F. is grateful to J.~Sanders for useful suggestions.
This work was partially supported by EPSRC grant No. GR/M61450 and by
DGES grant PB98-0821.

\section*{Appendix}

For the convenience of the reader we summarize
some known results of the theory
of master symmetries. We first review two additional
methods for constructing master symmetries which also
rely on the existence of a scaling. We then recall the relation
between master symmetries and time-dependent symmetries.\\

\ni{\bf Construction~1\/}~\cite{Fo87} {\em Let $\Phi$ be a\/ {\em hereditary
(Nijenhuis)} operator~\cite{Fu79}. Assume that
$\Phi$ is invariant under the scaling $\tau_0$
defined by Eq.~\eqref{tau0}. Then
$\tau=\Phi\tau_0$ is a master symmetry associated with
$$
\big\{q_x,K_1=\Phi q_x,K_2=\Phi^2 q_x\big\}\,.
$$}\vskip 1mm
For example, the hereditary operator associated with the KdV
equation~\eqref{potKdV} is
$$
\Phi=D_x^2+4q+2q_1\pa_x^{-1}\,.
$$
This operator is invariant under the
scaling $q\mapsto\al^{-2} q$, $x\mapsto\al\,x$,
which gives rise to $\tau_0=2q+xq_1$. Thus
$$
\tau=\Phi\tau_0=x(q_3+6qq_1)+4 q_2+8q^2+2q_1\pa_x^{-1}q
$$
is a master symmetry of the KdV equation.\\

\ni{\bf Construction~2\/}~\cite{Fu83} {\em Suppose that
$K_1(q),K_2(q)\in\cR\{q,\pa^{-1}q\}$ commute, and that $K_2$ admits a scaling
$\tau_0$. Assume that there exists a function $F\in\cR\{q,\pa^{-1}q,\bx\}$ such
that
\begin{equation}\label{F}
  \tau_0=[K_1\,,F]_L\,.
\end{equation}
Define $\tau(q)\in\cR\{q,\pa^{-1}q,\bx\}$ by
\begin{equation}\label{tauF}
  \tau=[K_2\,,F]_L\,.
\end{equation}
Then $\tau$ satisfies $[K_1\,,\tau]_L=c_2K_2$.}

Indeed,
$$
[K_1\,,\tau]_L=\big[K_1\,,[K_2\,,F]_L\big]_L=
\big[K_2\,,[K_1\,,F]_L\big]_L-\big[F\,,[K_1\,,K_2]_L\big]_L=
[K_2\,,\tau_0]_L=c_2K_2\,,
$$
where the fact that $\tau_0$ is a scaling of $K_2$ has been
used in the last equality.\\

\ni{\bf Time-dependent symmetries.} Let $\tau\in\cR\{q,\pa^{-1}q,\bx\}$
be a master symmetry associated with $\{q_x,K_1,K_2\}$,
$K_1,K_2\in\cR\{q,\pa^{-1}q\}$, normalized so that $[K_1\,,\tau]_L=K_2$.
Then
$$
\si=t K_2+\tau
$$
is a time-dependent generalized symmetry of the evolution equation
$q_t=K_1$.

\end{document}